\documentclass[a4paper,11pt]{article}
\usepackage{theorem}
\usepackage{latexsym}
\usepackage{amsmath}

\usepackage[dvips]{graphicx}
\usepackage{graphicx}

\newcommand{\newc}{\newcommand}

\newc{\be}{\begin{equation}}
\newc{\ee}{\end{equation}}
\newc{\bea}{\begin{eqnarray}}
\newc{\eea}{\end{eqnarray}}
\newc{\beas}{\begin{eqnarray*}}
\newc{\eeas}{\end{eqnarray*}}

\newc{\pardt}{\partial_{t}}
\newc{\pardxi}{\partial_{i}}
\newc{\pardts}{\partial_{t^{*}}}
\newc{\pardxis}{\partial_{i^{*}}}
\newc{\pardxj}{\partial_{j}}
\newc{\pardxk}{\partial_{k}}
\newc{\pard}{\partial}

\newc{\s }{\overline}

\newc{\sect}{\section}
\newc{\subs}{\subsection}

\newc{\defi}{\definition}
\newc{\prop}{\proposition}
\newc{\rem}{\remark}
\newc{\lem}{\lemma}
\newc{\exa}{\example}
\newc{\theo}{\theorem}
\newc{\coro}{\corollary}
\newc{\post}{\postulate}
\newc{\state}{\statement}

\begin{document}
\baselineskip0.5cm
\renewcommand {\theequation}{\thesection.\arabic{equation}}
\title{Phenomenological description of the vortex density in rotating BEC superfluids}

\author{M. Sciacca$^1$,
D.~Jou$^2$ 
and M.S.~Mongiov\`{\i}$^1$}

\date{}
\maketitle
\begin{center} {\footnotesize $^1$ Dipartimento di Metodi e Modelli Matematici
Universit\`a di Palermo, c/o Facolt\`{a} di Ingegneria,\\ Viale delle Scienze, 90128 Palermo, Italy\\
$^2$ Departament de F\'{\i}sica, Universitat Aut\`{o}noma de
Barcelona, 08193 Bellaterra, Catalonia, Spain}

PACS number(s): 67.40.Vs, 03.75.Lm
\end{center} \footnotetext{E-mail addresses: msciacca@unipa.it
(M. Sciacca), david.jou@uab.es (D. Jou), mongiovi@unipa.it (M. S.
Mongiov\`{\i})}

\begin{abstract}
We propose a phenomenological equation for the vortex line density
in rotating Bose-Einstein condensates as a function of the angular
speed. This equation provides a simple description of the gross
features of the increase in vortex number from the appearance of
the first vortex to the theoretical rigid-body result for high
vortex density, and allows one to compare with analogous
situations in superfluid helium, after the suitable changes in the
relevant parameters are made.

\end{abstract}

\section{Introduction} The study of quantized vorticity in
superfluid helium 4 has been a relevant topic in superfluidity for
several decades \cite{Do}. In the last decade, much interest has
been focused also on the study of quantized vortices in rotating
Bose-Einstein condensates of alkali-atomic gases
\cite{FS}--\cite{SR-PRA70},  for several geometries of the
confining potential. Here we focus our attention on the vortex
density $L$ per unit transversal area in rotating BEC in terms of
the angular velocity $\Omega$,  between the lower critical
velocity for the appearance of the first vortex to fast rotations,
where vortex discreteness may be neglected, and the system is
expected to be characterized by the rigid body result
$L=2\Omega/\kappa$, with $\kappa=h/m$ being the vorticity quantum,
$h$ Planck's constant and $m$ the mass of the particles.

To compare the behavior of vortices in both systems, helium 4 and
BEC, it is interesting to show macroscopic analogies and
differences arising from the respective microscopic features of
these systems, as for instance the low density, the long coherence
length, the size of the vortex radius, and the compressibility of
atomic BEC as compared with the high density, the extremely thin
vortices, and the incompressibility of liquid helium II.
Furthermore, due to very low temperatures, all particles of BEC
participate in the superfluid component, whereas in liquid helium
only a part of the particles are in the superfluid component and
the other ones in the normal viscous component. For the mentioned
comparison, it may be useful to explore phenomenological
frameworks common to both systems in which the macroscopic
behavior may be put in a common ground.

The aim of this paper is to adapt an evolution equation for the
vortex line density proposed for rotating superfluid helium
\cite{MJ1,MJ3} to the vortex density in BEC superfluids, by taking
into account the physical differences in the respective
situations. This is done in Section 2, whereas Section 3 is
devoted to the study of the solutions of the equation and of their
stability, from which the vortex density is obtained in terms of
the angular velocity and the corresponding vortex number is
calculated. In Section 4 we discuss a physical interpretation of
the macroscopical equation. In Section 5 we discuss a special
situation where the number of vortices seems to have a maximum
instead of being a monotonically increasing function of the
angular velocity.

\section{Evolution equation for vortex density}
Recently, two of us proposed a phenomenological equation for the
evolution of the vortex line density $L$ per unit volume in
superfluid helium 4 under rotation and counterflow \cite{MJ1},
\cite{MJ3}. In the case of pure rotation in a cylindrical
container, such an equation takes the form \cite{MJ3}

\be\label{deLsudt-gen} \frac{d L}{dt}=-\beta \kappa L^2+
\left[\alpha_2\sqrt{\kappa \Omega}-\alpha_3
\frac{\kappa}{d}\right]L^{3/2}- \left[\beta_1 \Omega-\beta_3
\frac{\sqrt{\kappa\Omega}}{d}+\alpha_4 \frac{\kappa}{d^2}\right]L,
\ee where $d$ is the diameter of the container, and $\alpha_i$ and
$\beta_i$ dimensionless phenomenological coefficients, whose
values are obtained by fitting experimental data.

Here, we aim to explore the application of a suitably modified
form of this equation to the description of the vortex density in
 rotating BEC in terms of the rotation frequency. The
consequences of equation (\ref{deLsudt-gen}) have been studied in
detail for superfluid $^4$He \cite{MJ3}, where the situation
without vortices $L = 0$ becomes unstable for values of $\Omega$
higher than a critical value
$\Omega_{1}=(\beta_3/2\beta_2)^2\kappa/d^2$, and the vortex
density $L$ increases for increasing $\Omega$ beyond the critical
value $\Omega_{1}$. For high values of $\Omega$, it follows from
(\ref{deLsudt-gen}) that the steady-state solution $L$ exhibits
the well-known behavior $L=2\Omega/\kappa$ \cite{Do}. Thus,
equation (\ref{deLsudt-gen}) describes the increase from one
single vortex to many vortices, and it is more suitable than
taking directly the assumption $L=2\Omega/\kappa$. Note that for
parallel vortices of the same height, $L$ is simply the number of
vortices per unit transversal area, i.e. $L=N/A$, with $A$ the
total transversal area of the system. Therefore, in this situation
the vortex line density per unit volume coincides with the vortex
line density per unit transversal area, to which we will pay
attention here for rotating BEC.

In contrast with superfluid helium, Bose-Einstein condensates are
not confined in material cylindrical containers, but in optical or
magnetic potential traps, to which one may add a magnetic rotation
in analogy with rotating container experiments for superfluid
helium. It is often assumed that the trap has the axially
symmetric form $V(r)=\frac{1}{2}m \Omega_t^2 r^2$ with $r$ the
transversal radius and $\Omega_t$ the frequency trap, which is a
constant characterizing the confining potential, or an ellipsoidal
form $V(x,y,z)=m(\omega_x^2 x^2+\omega_y^2 y^2+\omega_z^2 z^2)/2$,
where $\omega_x$, $\omega_y$ and $\omega_z$ are the oscillator
frequencies characterizing the potential of the trap in the three
spatial directions, and that the $X$ and $Y$ axes rotate with a
frequency $\Omega$, fixed by the frequency of variation of a
magnetic field along the $X$ and $Y$ axes \cite{HHHMF}.

Comparing with the rigid cylindrical containers in usual
experiments on rotating  $^4$He superfluids, in Bose-Einstein
condensates one has not a definite value for the diameter $d$ of
the container. Here we will take instead of it the characteristic
size of the trap, which controls the spatial extent of the
one-particle ground state of the harmonic potential and which is
called the oscillator length, $a_{osc}\approx
\left(\kappa/\omega_\bot\right)^{1/2}$. It is also useful to
recall, for future applications, that in axially symmetric traps
the stationary number particle density in terms of the radius in
the Thomas-Fermi approximation is \cite{LMS, SR-PRA70}
\be\label{n(r)=} n(r)=\frac{h}{\pi}\omega_\bot\left(\frac{N
a}{l_z}\right)^{1/2}\left[1-\frac{r^2}{R_{TF}^2}\right],\ee with
$N$ the total number of particles of the condensate, $a$ the
s-wave scattering length appearing in the Gross-Pitaevskii
equation \cite{FS}--\cite{SR-PRA70}, $l_z$ the average extent of
the condensate on the $z$ direction --- taken as the direction of
the axis of rotation ---, $\omega_\bot$ the average transversal
frequency of the trap, given by
$\omega_{\bot}^2=(\omega_x^2+\omega_y^2)/2$, or simply by
$\Omega_t$ in symmetric traps, and $R_{TF}$ is the Thomas-Fermi
radius of the condensate, given by \cite{LMS, SR-PRA70}
\be\label{RTF}
R_{TF}=2\left[\frac{Na}{l_z}\right]^{1/4}\left(\frac{\kappa}{2\pi
\omega_\perp}\right)^{1/2}.\ee To apply equation
(\ref{deLsudt-gen}) to rotating BEC, we identify $d$ as $a_{osc}$,
and (\ref{deLsudt-gen}) becomes \be\label{deLsudt-OmeSegn} \frac{d
L}{dt}=-\beta \kappa L^2+ \alpha_2  \sqrt{\kappa \omega_{\bot}}
\left[\sqrt{\overline{\Omega}}-\frac{\alpha_3}{\alpha_2}\right]L^{3/2}-\beta_1
\omega_{\bot}
\left[\overline{\Omega}-\frac{\beta_3}{\beta_1}\sqrt{\overline{\Omega}}+
\frac{\alpha_4}{\beta_1}\right]L, \ee where
$\overline{\Omega}=\Omega/\omega_{\bot}$ is the dimensionless
angular velocity. A further relevant difference with the usual
rotating cylinder used in superfluid helium may be the anisotropy
of the trap. In this case the coefficients of
(\ref{deLsudt-OmeSegn}) become a function of the eccentricity
$\epsilon$, defined as
$\epsilon=(\omega_y^2-\omega_x^2)/(\omega_x^2+\omega_y^2)$. We do
not pretend that equation (\ref{deLsudt-OmeSegn}) yields a full
explanation for the dynamics of the BEC, which would require to
include into the description other collective modes, but we aim to
identify some of the most salient phenomenological analogies and
differences with superfluid helium, which should, in the future,
to be understood from a microscopic basis of the Gross-Pitaevskii
equation.

\section{Stationary solutions of the evolution equation}
Now, we study the stationary solutions of equation
(\ref{deLsudt-OmeSegn}) and their corresponding domain of
stability. The stationary solutions are

\be\label{L=0-sol}
  L=0,
  \ee
  \be\label{L1/2-sol}
  L^{1/2}_\pm=\frac{\sqrt{\omega_\bot}}{2\beta \sqrt{\kappa}}
  \left(\alpha_2  \sqrt{\overline{\Omega}}-\alpha_3\right)\pm
  \sqrt{\frac{\omega_\bot}{4\kappa
  \beta^2}\left[\left(\alpha_2  \sqrt{\overline{\Omega}}-\alpha_3\right)^2
  -4\kappa \beta\beta_1  \left(\overline{\Omega}-\frac{\beta_3}{\beta_1}
  \sqrt{\overline{\Omega}}+\frac{\alpha_4}{\beta_1}\right)\right]}.
\ee

The first solution ($L=0$) corresponds to the absence of vortices,
and the second one describes an increase in the vortex density for
increasing $\Omega.$  Regarding the stability of solution $L = 0$
(no vortices), and according to equation (\ref{deLsudt-OmeSegn}),
we consider the evolution equation for the perturbation $\delta L$
around $L = 0$, which is

\be\label{deltaL} \frac{\delta L}{dt}=-\beta_1
\omega_{\bot}\left[\overline{\Omega}-\frac{\beta_3}{\beta_1}\sqrt{\overline{\Omega}}+\frac{\alpha_4}{\beta_1}\right]\delta
L. \ee From this equation we can establish that the solution $L=0$
is stable if

\be\label{beta1-magg0}
\beta_1\overline{\Omega}-\beta_3\sqrt{\overline{\Omega}}+\alpha_4\geq
0.\ee In superfluid helium \cite{MJ3}, a satisfactory choice of
the parameters is to assume $\beta_3^2=4\alpha_4\beta_1$, in which
case the critical value of $\s \Omega$ is \be\label{Omegabar_C}
{\s \Omega}_c
=\left(\frac{\beta_3}{2\beta_1}\right)^2=\frac{\alpha_4}{\beta_1},\ee
and $L=0$ is stable for $\s \Omega < \s \Omega_c.$ For symmetric
traps, the critical value in BEC is $\s \Omega_c=1/\sqrt{2},$
corresponding to half the frequency of the quadripole mode
\cite{MCBD}--\cite{TKU} which is a collective surface mode with
angular momentum $l=2$; when resonantly excited, quantized
vortices formed on the surface come into the condensate. In
eccentric traps, the value of $\s \Omega_c$ depends on the
eccentricity $\epsilon$, and it is close to $\s
\Omega(\epsilon)\cong \frac{1}{\sqrt{2}}-\epsilon$ (some authors
\cite{LMS} give this value whereas other ones \cite{HHHMF}
indicate $\s \Omega(\epsilon)\cong \frac{1}{\sqrt{2}}-0.91
\epsilon$). In any case, the critical value of $\s \Omega$ for the
formation of the first vortices is lower in eccentric traps than
in symmetric traps, which reflects the fact that the finite energy
barrier that vortices must overcome to move into the condensate is
lowered if the condensate is elongated.

For $\s \Omega \geq \s \Omega_c,$  the behavior of $L$ in terms of
$\s \Omega$ is described by (\ref{L1/2-sol}), which tends to $L=2
\s \Omega\omega_\bot/\kappa$ for high values of $\s \Omega.$

\begin{figure}[ht]
  \includegraphics[width=12cm]{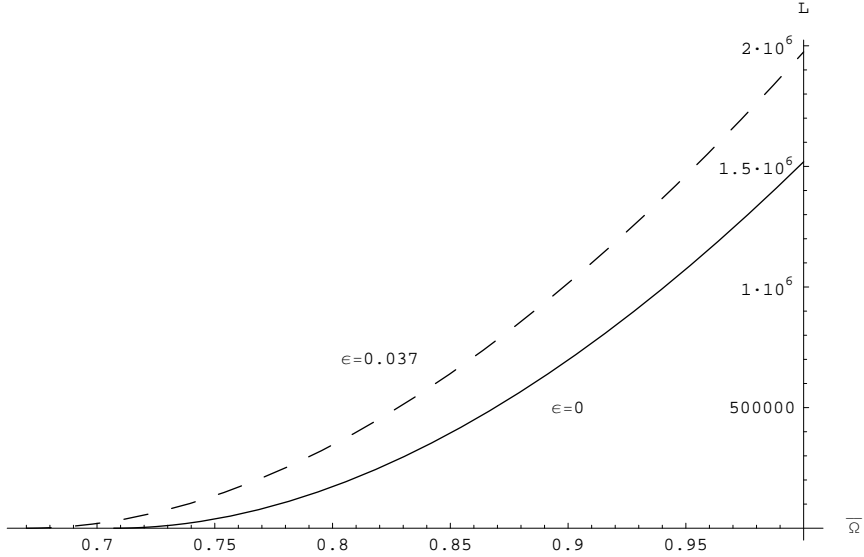}\\
  \caption{Vortex density $L$ in a BEC in terms of the dimensionless angular velocity
  for two values of the eccentricity $\epsilon = 0$ (continuous line) and $\epsilon = 0.037$ (dashed line).
  The graphics are plotted using
  the value $\omega_\bot /2\pi = 219\  \textrm{Hz}$ for the the average transversal frequency of the trap
  from Ref.~\cite{MCWD}.}\label{figura22}
\end{figure}

Under the previous assumption $\beta_3^2=4\alpha_4 \beta_1,$
equation (\ref{deLsudt-OmeSegn}) may be rewritten as
\be\label{deLsudt-OmeSegn2} \frac{d L}{dt}=-\beta \kappa L^2+
\alpha_2  \sqrt{\kappa \omega_{\bot}}
\left[\sqrt{\overline{\Omega}}-\sqrt{\overline{\Omega}_c}\right]L^{3/2}-\beta_1
\omega_{\bot}
\left[\sqrt{\overline{\Omega}}-\sqrt{\overline{\Omega}_c}\right]^2L,
\ee whose stable solution for $\s \Omega>\s \Omega_c$ is simply
\be\label{Lsoluzioni} L^{1/2}=\alpha_2
\frac{\sqrt{\omega_\bot}}{2\beta}\left[1+\sqrt{1-\frac{4\beta\beta_1
}{\alpha_2^2}}\right]
\frac{\sqrt{\overline{\Omega}}-\sqrt{\overline{\Omega}_c}}{\sqrt{\kappa}}=\sqrt{2
\omega_\bot}
\frac{\sqrt{\overline{\Omega}}-\sqrt{\overline{\Omega}_c}}{\sqrt{\kappa}}.\ee
The value of the combination of coefficients appearing in the
prefactor of (\ref{Lsoluzioni}) is dictated by the fact that for
$\s \Omega\gg\s \Omega_c$, $L$ tends to the rigid body result $L
\cong 2 \s \Omega \omega_\bot/\kappa$. In Fig.~\ref{figura22} we
plot $L$ in terms of $\s \Omega$ for a given $\omega_\bot$ and for
two values of eccentricity of the potential, $\epsilon = 0$
(continuous line) and $\epsilon = 0.037$ (dashed line)
\cite{MCWD}. As expected, graphics of $L$ in figure~\ref{figura22}
confirm the fact that when the eccentricity increases the first
vortex in BEC appears for a smaller value of rotation. Expression
(\ref{Lsoluzioni}) generalizes the well-known expression
$L=2\Omega/\kappa$ and, from a practical perspective, it is more
useful than it, because in BEC, in contrast to helium 4, the
angular velocity cannot be increased indefinitely, because of the
effects of the centrifugal force. Note, of course, that these
lines are not completely realistic, and only discontinuous values
of them, corresponding to integer numbers of vortices, are
actually meaningful.
\begin{figure}[ht]
  \includegraphics[width=12cm]{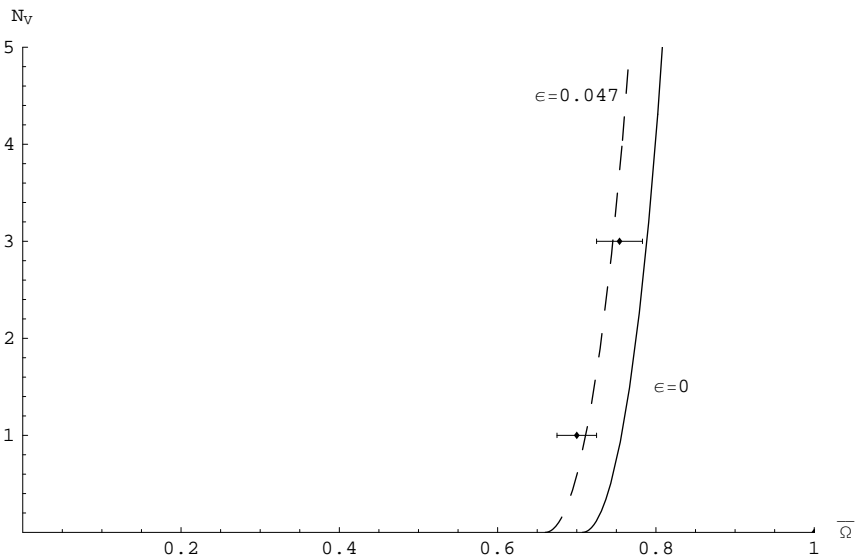}\\
  \caption{The number of vortex $N_V$ in a BEC in terms of the dimensionless angular velocity for
   two values of the eccentricity $\epsilon = 0$ (continuous line) and $\epsilon = 0.037$ (dashed line).
   The graphics are plotted using the values $\omega_\bot /2\pi =219\  \textrm{Hz}$, $N=10^5$, $a=8\ \mu \textrm{m}$ and
  $l_z=49\ \mu \textrm{m}$ from Ref.~\cite{MCWD}. The horizontal lines
 indicates the ranges of $\s \Omega$  to which  the number of vortices corresponds: the first
 range corresponds to $1$ vortex while the second one corresponds
 to $2/4$ vortices.}\label{figura11}
\end{figure}
\par
In many analyses, one measures the number $N_V$ of the vortices
rather than their areal density. To go from $L$ to $N_V$ one must
multiply $L$ times the transversal area of the condensate, taking
into account that the value of its radius depends on the rotation
rate, because of the centrifugal force \cite{SR-PRA70}. One has
\be\label{N_V} N_V=L\pi R^2=\frac{L\pi
R_{TF}^2}{1-(\Omega/\omega_\bot)^2}=\frac{L\pi R_{TF}^2}{1-\s
\Omega^2},\ee with $R_{TF}$ the Thomas-Fermi radius (\ref{RTF}).
Note, then, that for a given value of the angular frequency the
total number of vortices increases when the number of particles
increases, proportionally to $(N a/l_z)^{1/2}$; thus, it depends
not directly on $N$, but rather on the number of particles per
unit transverse length. A second aspect is that $\Omega$ cannot
increase indefinitely, because at $\Omega=\omega_\bot$ the
centrifugal force is so high that the condensate is flattened out
and it may disperse, unless a suitable quartic potential is added
to the harmonic trap. However, since both $L$ and $R$ increase
with $\Omega,$ it follows that as far as the condensate exists,
the total number of vortices given in (\ref{N_V})  increase as
function of the angular velocity, as shown in Fig.~\ref{figura11},
where $N_V$ is plotted for two values of the eccentricity
$\epsilon = 0$ (continuous line) and $\epsilon = 0.037$ (dashed
line). There a direct comparison between the graphics of our model
and the experimental data of Ref.~\cite{MCWD} is carried out.
Note, incidentally, that the relation between $L$ and $N_V$ is
very different in BEC than in confined superfluid helium, where
the number of vortex is
\be\label{N_Vhelium}%
N=L \pi R^2\approx\frac{2 \pi R^2
\Omega}{\kappa}\left[1-\sqrt{\frac{\Omega_c}{\Omega}}\right].
\ee%
with $R$ the radius of the cylinder, $\kappa$ the vorticity
quantum --- which is different with respect to that of BEC --- and
$\Omega_c$ the critical rotation velocity for the appearing of the
first vortex in rotating helium II.
  The difference between two relations ((\ref{N_V}) and (\ref{N_Vhelium})) is caused by the
fact that in the latter the radius of the sample always coincides
with the radius of the container and it does not depend on the
angular velocity, whereas in BEC the change in the radius with
$\Omega$ is very relevant.

To compare the behavior of vortices in rotating Bose-Einstein
condensates and in rotating superfluid helium 4, it is also worth
to stress that the characteristic values of the angular velocities
and of the vortex densities are rather different in both cases; in
typical experiments with helium the angular frequency is less than
$10 \ \textrm{rad/s}$, whereas in BEC may be of the order of $300\
\textrm{rad/s}$; and the vortex line densities are of the order of
$10^4\ \textrm{vortex lines}/\textrm{cm}^2$ for superfluid helium
and of $10^7\ \textrm{vortex lines}/\textrm{cm}^2$ in BEC.
However, we should not compare the actual values of $\Omega$ but
characteristic dimensionless rotation velocity. In BEC, it is
defined as $\overline{\Omega}\equiv\Omega/\omega_{\bot}$, whereas
for superfluid helium it is given by $\overline{\Omega}\equiv
\Omega d^2/\kappa$; for $^4\textrm{He}$, the value of the quantum
of vorticity $\kappa$ is $\kappa \approx 9.97\times 10^{-4}
\textrm{cm}^2 \textrm{s}^{-1}$; thus, the value of the
dimensionless velocity corresponding to $10\
\textrm{rad}/\textrm{s}$ for helium in a container of diameter $1
\textrm{cm}$, would be of the order of $10^5$, much higher than
those explored in Bose-Einstein condensates.  This difference  can
also  be related to the fact that the quanta of vorticity of
helium II and BEC are very different, as the quantum of vorticity
in rubidium BEC is about $20$ times lower than in helium II
because of the lower atomic mass of helium with respect to that of
rubidium.

The equation (\ref{deLsudt-OmeSegn}) has been applied to
superfluid helium in rotating cylinders. It would be interesting
to have information for rotating containers with ellipsoidal cross
section, in order to explore the influence of the eccentricity on
the vortex line density, but we are not aware of experimental
results in this problem. A parallel possibility would be to
compare rotating containers with rectangular cross section
\cite{YG}.

\section{A physical interpretation of the evolution equation}
To have a more explicit understanding of equation
(\ref{deLsudt-OmeSegn}) or (\ref{deLsudt-OmeSegn2}) we recall that
in rotating Bose-Einstein condensates and Helium II, the vortices
are produced at the surface of the condensate and penetrate into
it pulled by the rotating drive. The repulsive interaction amongst
them tends to push them apart because vortices rotating in the
same direction experience an effective repulsive interaction
\cite{MCWD, MCBD}. This competition between centripetal driving
and repulsive force yields eventually to a regular vortex lattice.
This idea suggests us to identify the terms appearing in equation
(\ref{deLsudt-OmeSegn}) in terms of these features. In this way,
we propose to interpret (\ref{deLsudt-OmeSegn}) as

\be\label{deLsudt-in-out} \frac{d L}{d t}=[v_{in}-v_{out}]L^{3/2},
\ee with $v_{in}$, $v_{out}$ being the respective ingoing and
outgoing vortex drift velocities, proportional to the respective
driving forces. The exponent $3/2$ for $L$ in
(\ref{deLsudt-in-out}) is chosen on dimensional grounds, since $L$
has dimensions of $(\textrm{lenght})^{-2}$. There is not an
univocal way to decide to which term, either $v_{in}$ or
$v_{out}$, must be assigned each term in (\ref{deLsudt-OmeSegn})
or in or (\ref{deLsudt-OmeSegn2}). Here we propose to relate the
terms linear in $L^{3/2}$ to the inflowing flux, because the
corresponding velocity would not depend on the vortex density, but
only on the rotation. Following this tentative criterion, the
respective values of $v_{in}$ and $v_{out}$ would be
\be\label{v_out} v_{out}=\beta\kappa
L^{1/2}+\beta_1\omega_\bot\left[\sqrt{\s \Omega}-\sqrt{\s
\Omega_c}\right]^2L^{-1/2}, \ee

\be\label{v_in} v_{in}=\alpha_2\sqrt{\kappa
\omega_\bot}\left[\sqrt{\s \Omega}-\sqrt{\s \Omega_c}\right]. \ee

The outgoing velocity depends on the vortex line density: it
depends on the inverse of the average vortex separation (which is
given by $L^{-1/2}$). Thus, (\ref{v_out}) could be considered as
the first terms in a kind of virial expansion of a repulsive force
between vortices in terms of powers of $L^{1/2}$. In vortex arrays
in rotating containers, the free energy has a term which may be
interpreted as a repulsion force. The ingoing velocity is related
to the angular velocity: it is positive at values of
$\overline{\Omega}$ higher than $\overline{\Omega}_c$. This could
probably be related to an energy barrier which the vortices must
overcome in order to move into the condensate. Then, the
interpretation of a dynamical equation for $L$ for rotating
superfluids --- BEC or liquid helium
--- is rather different than for counterflow systems, where it
relies on the dynamics of vortex breaking and reconnections, as
proposed Schwarz \cite{S1}, whose microscopic view is not useful
for parallel vortex lines. Furthermore, in Schwarz microscopical
model for the dynamics of $L$ under counterflow in superfluid
helium, all the volume contributes to the production term of $L$,
because of the elongation of vortex loops, whereas in rotating
systems the vortices are produced on the surface. A relevant
difference between BEC and superfluid helium may be the extent of
vortex interaction, because of the considerable width of vortices
in BEC as compared with the atomic width of vortex lines in
superfluid helium. This effect should be reflected in the value of
coefficient $\beta$.

\section{An anomalous non-monotonic density variation}
The usually expected behavior of $L$ in terms of $\Omega$ is a
monotonical increase as mentioned in Section 3. However, in an
experiment by Hodby {\it et al.} \cite{HHHMF} a different behavior
was observed, which we briefly report here for the sake of
completeness, although it seems rather exceptional, as it has not
been reproduced, to our knowledge. In \cite{HHHMF} Hodby {\it et
al.} studied experimentally the evolution of the number of
vortices for a condensate of $2\cdot 10^4$ atoms of
$^{87}\textrm{Rb}$ for an oblate geometry ($\omega_{\bot}<
\omega_z$) within a rotating magnetic elliptical trap. They
observed that when the eccentricity is adiabatically ramped from
$\epsilon = 0$ to a given final value $\epsilon$, there were both
a lower $\overline{\Omega}_1(\epsilon)$ and an upper
$\overline{\Omega}_2(\epsilon)$ values of rotation rates for which
vortices were nucleated; these values depend on the eccentricity,
and were given respectively by \cite{HHHMF}

\be\label{Omega1} \overline{\Omega}_1(\epsilon)\approx0.71-0.91
\epsilon, \ee and

\be\label{Omega2}
\frac{2}{\overline{\Omega}_2}\left[\frac{2\overline{\Omega}_2^2-1}{3}\right]^{2/3}=\epsilon.
\ee

Hodby {\it et al.} \cite{HHHMF} measured the number of vortices as
a function of  $\overline{\Omega}$ for two values of the
eccentricity, and found that this has a maximum ($L_{max}$,
$\overline{\Omega}_{max}$) within the interval
$\overline{\Omega}_1 \leq \overline{\Omega}\leq
\overline{\Omega}_2$. This information may also be synthesized
into (\ref{deLsudt-OmeSegn}), by identifying its coefficients in
terms of $\overline{\Omega}_1(\epsilon)$,
$\overline{\Omega}_2(\epsilon)$,
$\overline{\Omega}_{max}(\epsilon)$ and $L_{max}(\epsilon)$.

To describe Hodby's results one should assume $\beta_3^2>4
\alpha_4\beta_1 $ instead of the equality of both terms assumed in
(\ref{Omegabar_C}). In this case, there will be an instability
range for $L=0$ for $\overline{\Omega}$ between
$\overline{\Omega}_1^{1/2}$ and $\overline{\Omega}_2^{1/2}$ given
by

\be\label{Omega1^1/2}
\overline{\Omega}_1^{1/2}=\frac{\beta_3}{2\beta_1}-\frac{\sqrt{\beta_3^2-4
\alpha_4 \beta_1}}{2\beta_1}, \ee

\be\label{Omega2^1/2}
\overline{\Omega}_2^{1/2}=\frac{\beta_3}{2\beta_1}+\frac{\sqrt{\beta_3^2-4
\alpha_4 \beta_1}}{2\beta_1}, \ee from which we deduce immediately
that
$\beta_3/\beta_1=\sqrt{\overline{\Omega}_1}+\sqrt{\overline{\Omega}_2}$
and $\alpha_4/\beta_1=\sqrt{\overline{\Omega}_1
\overline{\Omega}_2}$.

Now, we rewrite the non zero stationary solution (\ref{L1/2-sol})
as

\be\label{radL-pm} \sqrt{L_{\pm}}=\frac{\alpha_2
\sqrt{\omega_{\bot}}}{2 \sqrt{\kappa}}
\left(\sqrt{\overline{\Omega}}-\sqrt{\overline{\Omega}_2}\right)
\left[1\pm \sqrt{1+\frac{4\beta_1 }{\alpha_2^{\prime2}}
\frac{\sqrt{\overline{\Omega}}-\sqrt{\overline{\Omega}_1}}{\sqrt{\overline{\Omega}_2}-\sqrt{\overline{\Omega}}}}\right],
\ee with $\alpha_2 =\alpha_2/\beta$ and $\beta_1 =\beta_1/\beta$.
We have taken $\alpha_3/\alpha_2=\sqrt{\overline{\Omega}_2}$ in
order to ensure that the solution of $L$ will vanish at
$\overline{\Omega}=\overline{\Omega}_2$. This condition implies a
restriction on the coefficients in (\ref{deLsudt-OmeSegn}), namely
$\alpha_3/\alpha_2= \beta_3/\beta_1-(\alpha_4 \alpha_2)/(\beta_1
\alpha_3).$

The branch $L_+$  is not considered here because inside the
interval $\left[\overline{\Omega}_1,\overline{\Omega}_2\right]$ it
is negative; whereas $L_-$ vanishes at $\overline{\Omega}_1$ and
$\overline{\Omega}_2$ and it has a maximum for a value of the
frequency $\sqrt{\overline{\Omega}}_{max}$, given by the condition

\be\label{alfa2/rad-Condiz} \frac{\alpha _2}{\sqrt{\beta _1
}}=\frac{\sqrt{\overline{\Omega}_{max}}-\sqrt{\overline{\Omega}_1}}{\sqrt{\overline{\Omega}_2}-\sqrt{\overline{\Omega}_{max}}}-1.\ee

The corresponding maximum value $L_{max}$ is

\be\label{beta'1=} \beta _1=\frac{\kappa L_{max}}{\omega_{\bot}
\left[\sqrt{\overline{\Omega}_2}-\sqrt{\overline{\Omega}_{max}}\right]^2}.
\ee

Thus, the coefficients in (\ref{deLsudt-OmeSegn}) have been
identified in terms of $\overline{\Omega}_1(\epsilon)$,
$\overline{\Omega}_2(\epsilon)$,
$\overline{\Omega}_{max}(\epsilon)$ and $L_{max}(\epsilon)$, from
which their values could be found,  except for $\beta$, related to
the dynamics of the vortex line density. The coefficients
$\alpha_i$, $\beta_i$ depend on the eccentricity $\epsilon$, but
not on $\omega_\bot$ neither on the mass of the particles, which
appears in $\kappa$.

To compare with the experimental values of Hodby {\it et al.}
\cite{HHHMF} one must go from the values of $L$ (number of
vortices per unit area) to the average number of vortices in the
condensate, $N$, which are related through  $L=N/A$, with $A$ the
transversal area of the condensate. This will be estimated as
$A=(\pi/4)d^2=(\pi/4)(\kappa/\omega_{\bot})$. In fact, it must be
recalled that the transverse radius depends on the angular
velocity $\Omega$ as commented in (\ref{N_V}). In the trap used by
Hodby {\it et al.} \cite{HHHMF}, $\omega_{\bot}\approx124 \pi
\textrm{s}^{-1}$, and for $^{87}\textrm{Rb}$, the vorticity
quantum $\kappa=h/m$ is $4.54\times 10^{-5} \textrm{cm}^2
\textrm{s}^{-1}$. Thus, we take for the average area of the
condensate $A\approx 9.15\times 10^{-8}
\textrm{cm}^2$. In Fig. 3 
the curves of our model and
the experimental results by Hodby {\it et al.} are shown
\cite{HHHMF}. The experimental data are disperse but the general
qualitative features are well exhibited. We have taken for
$\overline{\Omega}_1$, $\overline{\Omega}_2$,
$\overline{\Omega}_{max}$ and $L_{max}$ their experimental values
--- in fact, the experimental value for $\overline{\Omega}_2$ found by Hodby
{\it et al.} does not coincide exactly with the theoretical value
coming from (\ref{Omega2}).

Hodby {\it et al.} give the value of $\overline{\Omega}_{max}$  for
two values of  $\epsilon$, namely, $\epsilon=0.041$, for which
$\overline{\Omega}_{max}\approx 0.74$, and  $\epsilon=0.084$ for
which $\overline{\Omega}_{max}\approx 0.785$. In view of the values
reported by these authors, we suggest for
$\overline{\Omega}_{max}(\epsilon)$ the ansatz

\be\label{Omega_max} \overline{\Omega}_{max}(\epsilon)\approx
\frac{\overline{\Omega}_c+\overline{\Omega}_2(\epsilon)}{2}=\frac{\sqrt{2}}{4}+
\frac{1}{2}\overline{\Omega}_2(\epsilon).\ee In fact, the values
of $\overline{\Omega}_2(\epsilon)$  observed in Fig. 1 of
\cite{HHHMF} were $\overline{\Omega}_2(\epsilon=0.041)\approx
0.77\pm 0.02$ and $\overline{\Omega}_2(\epsilon=0.084)\approx
0.86\pm 0.02$, in such a way that (\ref{Omega_max}) yields indeed
a reasonable estimate for $\overline{\Omega}_{max}(\epsilon)$ in
these cases.

\section{Conclusions}
We adapted a previous dynamical equation proposed for the vortex
density in superfluid helium to BEC condensates. The adaptation is
not obvious, because it requires to consider the form of the trap
potential, the radius of the BEC cloud in terms of the number of
particles and the angular velocity, and information on the
critical value of $\Omega$ related to the resonant excitation of
the quadrupolar mode.

An interesting aspect of our proposal is the simplicity of the
expression (\ref{Lsoluzioni}) for the vortex density in term of
the angular rotation. The lines shown in the figure are
indicative, as in fact the number of vortices only take
discontinuous values, whereas the actual values are related to the
discrete number of vortices. Relation (\ref{Lsoluzioni}) shows the
considerable difference between the vortex density starting from
$L=0$ at $\Omega=\frac{\sqrt{2}}{2}\omega_\bot$ and tending to
$L=\frac{2\Omega}{\kappa}$ for an asymptotic limit of $\Omega\gg
\omega_\bot$ which is never reached, as the centrifugal force
limits $\Omega$ to values less or equal than $\omega_\bot$, with
$\omega_\bot$ the transversal frequencies of the confining trap.
Thus, (\ref{Lsoluzioni}) albeit phenomenological, may be useful
for simple estimates of the density, and also of the number of
vortices, as given by (\ref{N_V}).

We also included in Section 5 a seemingly anomalous situation
where the number of vortices is not an increasing function of
$\Omega,$ but is reported to have a maximum. To our knowledge,
this observation has not been replicated by other groups, and it
could be related to some metastable situation. In any case, we
have reported it here to have a better appreciation of the role of
some restrictions on the values of the numerical coefficients, as
that reported in (\ref{Omegabar_C}), whose breakdown could lead to
a situation as that reported in Section 5 instead to the much more
generic results of Section 3.

A physical interpretation of the meaning of the several terms in
(\ref{deLsudt-OmeSegn}) is still open, and is related to the
entrance into the condensate of vortices formed in the walls, and
their subsequent mutual interactions inside the condensate.
\\
\\
\\
\section*{Acknowledgments}
We acknowledge the support of the Acci\'{o}n Integrada
Espa\~{n}a-Italia (Grant S2800082F HI2004-0316 of the Spanish
Ministry of Science and Technology and grant IT2253 of the Italian
MIUR). DJ acknowledges the financial support from the
Direcci\'{o}n General de Investigaci\'{o}n of the Spanish Ministry
of Education under grant FIS 2006-12296-C02-01 and of the
Direcci\'{o} General de Recerca of the Generalitat of Catalonia,
under grant 2005 SGR-00087. MSM and MS acknowledge the financial
support from MIUR under  grant "PRIN 2005 17439-003" and by "Fondi
60\%" of the University of Palermo. MS acknowledges the "Assegno
di ricerca: Studio della turbolenza superfluida e della sua
evoluzione" of the University of Palermo.

\begin{figure}[h]
  \includegraphics[width=12cm]{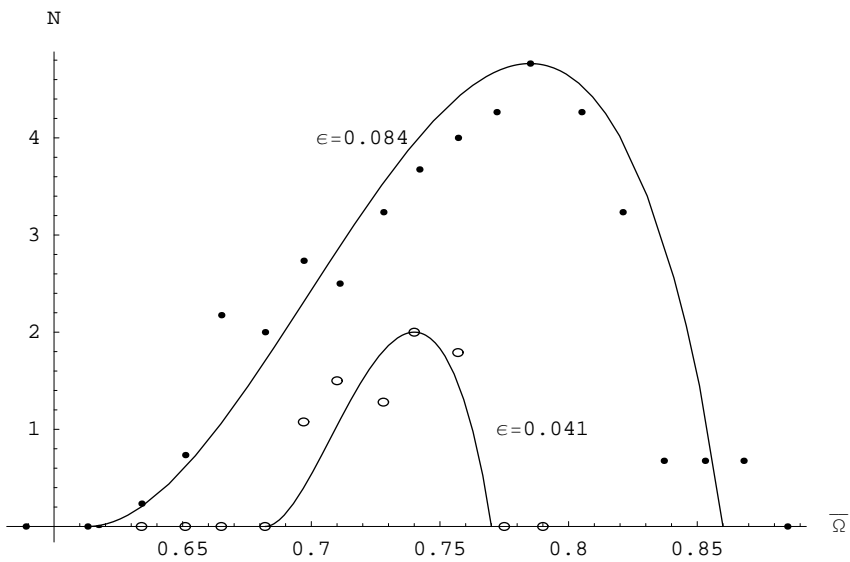}\\
  \caption{Representation of the number of vortices $N$ in terms of the dimensionless rotational velocity
  $\overline{\Omega}$ according to equation (\ref{radL-pm}) . The points correspond to the experimental
  data of Hodby {\it et al.} \cite{HHHMF}. The continuous lines
  correspond to our model, taking for $\overline{\Omega}_1$,
  $\overline{\Omega}_2$, $\overline{\Omega}_{max}$ and $L_{max}$ the
  experimental values. The upper line corresponds to the eccentricity
  $\epsilon_1=0.084$ and the lower one to  eccentricity
  $\epsilon_2=0.041$.
  }\label{figura1}
\end{figure}
\end{document}